\documentstyle[preprint,aps,psfig]{revtex}
\newcommand{\T}{{\large$t$}}
\begin{document}
\draft
\preprint{MKPH-T-96-6}
\title{The position and the residues of the delta resonance pole\\
 in pion photoproduction}

\author{O.~Hanstein, D.~Drechsel and L.~Tiator}
\address{Institut f\"ur Kernphysik, Universit\"at Mainz, 55099 Mainz, Germany}
%\date{\today}
\maketitle

\begin{abstract}
We have analyzed the $M_{1+}^{(3/2)}$ and $E_{1+}^{(3/2)}$ multipole
amplitudes of
pion photoproduction in the framework of fixed-$t$ dispersion relations.
Applying the speed plot technique to our results for these multipoles, we have
determined the position and the residues of the $\Delta$ (1232) resonance
pole. The pole is found at total $c.m.$ energy $W = (1211 - 50i)$ MeV on
the second Riemann sheet, and the ratio of the electric and magnetic
residues is $R_{\Delta} = - 0.035 - 0.046 i$,
resulting in an E2/M1 ratio for the "dressed" delta resonance of $- 3.5 \%$.

\end{abstract}

\section{Introduction}

The determination of the quadrupole excitation strength $E_{1+}^{(3/2)}$
in the region of the $\Delta (1232)$ resonance has been the aim of considerable
experimental and theoretical activities. Within the harmonic oscillator quark
model, the $\Delta$ and the nucleon are both members of the symmetrical
56-plet of $SU(6)$ with orbital momentum $L = 0$, positive
parity and a Gaussian wave function in space. In this approximation
the $\Delta$ may only be excited by a magnetic dipole transition
$M_{1+}^{(3/2)}$ \cite{Bec65}. However, in analogy with the atomic
hyperfine interaction or the
forces between nucleons, also the interactions between the quarks contain a
tensor component due to the exchange of gluons. This hyperfine interaction
admixes higher states to the nucleon and $\Delta$ wave functions, in
particular $d$-state components with $L = 2$, resulting in a small
electric quadrupole transition $E_{1+}^{(3/2)}$ between nucleon and $\Delta$
\cite{Kon80,Ger81,Isg82},
and a quadrupole moment of the $\Delta$ $Q_{\Delta} \approx -.09 fm^2$
\cite{Dre84}.
Therefore an accurate measurement of $E_{1+}^{(3/2)}$ is of great importance
in testing the forces between the quarks and, quite generally,
models of nucleons and isobars.

The ratio
\begin{equation}
\label{R_EM}
R_{EM} = \frac{\mbox{Re}\left[E_{1+}^{(3/2)}M_{1+}^{(3/2)\ast}\right]}
               {\mid M_{1+}^{(3/2)}\mid^{2}}
\end{equation}
 has been predicted to be
in the range  $- 2\%\le R_{EM} < 0\%$ in the framework of constituent quark
\cite{Kon80,Isg82,Dre84}, relativized
quark \cite{Bie87,Cap90,Cap92} and chiral bag models \cite{Kae83,Ber88}.
Considerably larger values have been obtained in
Skyrme mo\-dels \cite{Wir87}.
A first lattice QCD calculation resulted in a small value with large error
bars $(- 6 \% \le R_{EM} \le 12\%)$ \cite{Lei92}.
However,
the connection of the model calculations with the experimental data is not
evident. Clearly, the $\Delta$ resonance is coupled to the pion-nucleon
continuum and final-state interactions will lead to strong background terms
seen in the experimental data, particularly in case of the small $E_{1+}$
amplitude. The question of how to ''correct'' the experimental data to
extract the properties of the re\-so\-nance has been the topic of
many theoretical
investigations. As some typical examples we refer the reader to the work
of Olsson \cite{Ols74}, Koch et al. \cite{Koc84} and Laget \cite{Lag88}.
Unfortunately it turns out that the
analysis of the small $E_{1+}$ amplitude is quite sensitive to details of the
models, e.g. nonrelativistic vs. relativistic  resonance denominators,
constant or energy-dependent widths and masses of the resonance, sizes of
the form factor included in the width etc. In other words, by changing
these definitions
the meaning of resonance vs. background changes, too. More recently, Nozawa
et al. \cite{Noz90} have included final-state interactions in a
dynamical model with quark
cores and pions, and Davidson et al. \cite{Dav86} have analyzed
photoproduction in terms
of effective Lagrangians, taking account of final-state interactions
implicitly through unitarization. By fitting the parameters of the models to
older sets of data, Ref.~\cite{Noz90} obtained a ratio
$R_{EM}^{\mbox{\scriptsize bare }\Delta} = - 3.1 \%$ for the
bare $\gamma N\Delta$ coupling, while Ref.~\cite{Dav86} deduced a value of
$R_{EM}^{\mbox{\scriptsize res}} = - 1.4 \%$, including
some ''dressing'' from final-state interaction. A detailed discussion of
these models and a comparison to the data is given in Ref.~\cite{Kha93}. In an
extension of the work of Nozawa et al., Bernstein et al. \cite{Ber93}
have decomposed
the multipole contributions into resonant and background terms, and
compared their analysis to
previous investigations. As a result they obtained
$R_{EM}^{\mbox{\scriptsize bare }\Delta} = -(3.1 \pm 1.3)\%$
for the ''bare $\Delta$'' amplitude and
$R_{EM}^{\mbox{\scriptsize res}} = - 2.2\%$ for the
''dressed $\Delta$''. In very recent relativistic and unitarized pion
photoproduction calculations, for the ''bare $\Delta$'' ratios of
$R_{EM}^{\mbox{\scriptsize bare }\Delta} = -1.43\%$ \cite{Van95} and
$R_{EM}^{\mbox{\scriptsize bare }\Delta} = -1.46\%$ \cite{Sur95} are found.

In order to study the $\Delta$ deformation, 
pion photoproduction on the proton has recently been measured by the LEGS
collaboration \cite{Bla92} at Brookhaven and by the A2 collaboration  
\cite{Bec95} at MAMI using transversely
polarized photons, i.e. by measuring the polarized photon asymmetry
$\Sigma$. In particular, the cross section $d\sigma_{\parallel}$ for photon
polarization in the reaction plane turns out to be very sensitive to the small
$E_{1+}$ amplitude. Assuming for simplicity that only the $P$-wave multipoles
contribute, the differential cross section is
\begin{equation}
\frac{d\sigma_{\parallel}}{d\Omega} = \frac{q}{k} (A_{\parallel} + B_{\parallel}
\cos \Theta_{\pi} + C_{\parallel} \cos^2 \Theta_{\pi}),
\end{equation}
where $q$ and $k$ are the pion and photon momenta and $\Theta_{\pi}$ is
the pion emission angle in the $c.m.$ frame. Neglecting the (small)
contributions of the Roper multipole $M_{1-}$, one obtains \cite{Bec95}
\begin{equation}
C_{\parallel}/A_{\parallel} \approx 12 R_{EM} ,
\end{equation}
because the isospin $\frac{3}{2}$ amplitudes strongly dominate the cross
section for $\pi^{0}$ production.
In the meantime new precision data have been obtained by the A2
collaboration at MAMI with polarized photons for
both charged and neutral pion production over the energy range
270 MeV $\le E_{\gamma} \le$ 420 MeV \cite{Bec95,Kra96}. These data
will make it possible
to determine the partial wave amplitudes over the full region of the $\Delta$
resonance. The preliminary data for $\pi^{0}$ production are in good 
agreement with the ratio
$d\sigma_{\parallel} / d\sigma_{\perp}$ measured by the LEGS
collaboration, at the lower energies.

\section{Dispersion relations at fixed \T}

\begin{figure}[htbp]
\centerline{\psfig{figure=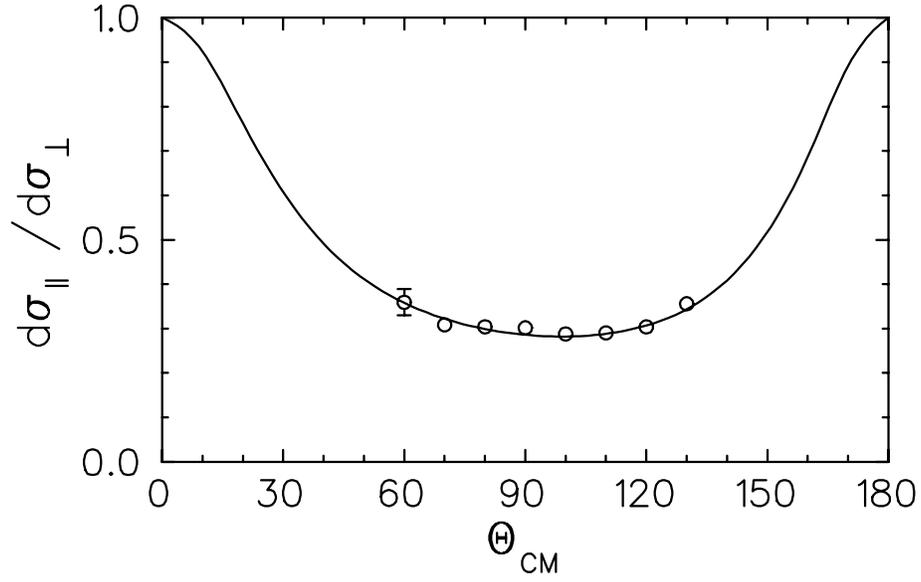,width=12cm}}
\vspace{0.5cm}
    \caption{Ratio $d\sigma_{\parallel}/d\sigma_{\perp}$ of the differential
     cross section for p($\gamma,\pi^{0}$)p at photon {\em lab.} energy
     $E_{\gamma}=333$~MeV. The solid line is our result from dispersion
     analysis. The data are from \protect\cite{San96}.}
  \label{fig:legs}
\end{figure}

\begin{figure}[htbp]
\centerline{\psfig{figure=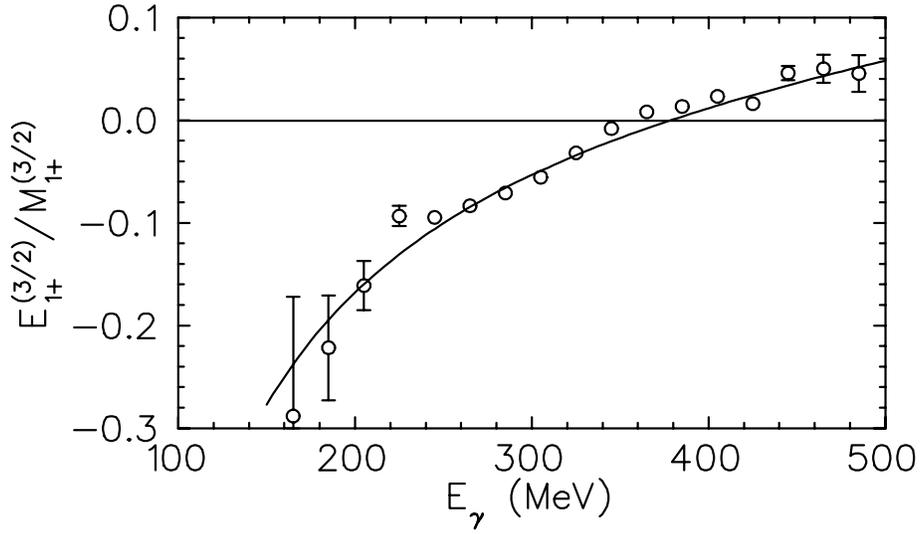,width=12cm}}
\vspace{0.5cm}
    \caption{Ratio of the electric and magnetic 3,3-multipoles as a function
     of the photon {\em lab.} energy. The solid line is the result of our
     dispersion analysis, the data points are from the
     VPI analysis \protect\cite{Arn95}.}
  \label{fig:e_over_m}
\end{figure}

Starting from fixed-$t$ dispersion relations for the invariant amplitudes
 of pion photoproduction, the projection of the multipole amplitudes leads
to a well known system of integral equations,
\begin{equation} \label{inteq}
\mbox{Re}{\cal M}_{l}(W) = {\cal M}_{l}^{\mbox{\scriptsize P}}(W)
+ \frac{1}{\pi}\sum_{l'}{\cal P}\int_{W_{\mbox{\scriptsize thr}}}^{\infty}
K_{ll'}(W,W')\mbox{Im}{\cal M}_{l'}(W')dW',
\end{equation}
where ${\cal M}_l$ stands for any of the multipoles
$E_{l\pm}, M_{l\pm},$ and ${\cal M}_{l}^{\mbox{\scriptsize P}}$ for the
corresponding (nucleon) pole
term. The kernels $K_{ll'}$ are known, and the real and imaginary parts of the
amplitudes are related by unitarity. In the energy region below two-pion
threshold, unitarity is expressed by the final state theorem of
Watson \cite{Wat54},
\begin{equation} \label{watson}
\cal{M}_{l}^{I} (W) = \mid \cal{M}_{l}^{I} (W)\mid e^{i(\delta_{l}^{I} (W)
+ n\pi)},
\end{equation}
where $\delta_{l}^{I}$ is the corresponding $\pi N$ phase shift and $n$ an
integer. We have essentially followed the method of Schwela et al
\cite{Sch69,Pfe72} to solve
Eqs. (\ref{inteq}) with the constraint (\ref{watson}). In addition
we have taken account of the
coupling to some higher states neglected in that earlier reference. At the
energies above two-pion threshold up to $W = 2$ GeV, Eq.~(\ref{watson}) has
been replaced by an ansatz based on unitarity \cite{Sch69}. Finally, the
contribution of the
dispersive integrals from $2$ GeV to infinity has been replaced by $t$-channel
exchange, parametrized by certain fractions of $\rho$- and $\omega$-exchange.
Furthermore, we have to allow for the addition of solutions of the
homogeneous equations to the coupled system of Eq.~(\ref{inteq}). The whole
procedure introduces
9 free parameters, which have to be determined by a fit to the data. In our
 data base we have included the recent MAMI experiments
for $\pi^{\circ}$ and $\pi^{+}$ production off the proton in the
energy range from 160  MeV to 420 MeV \cite{Fuc96,Kra96,Hae96},
both older and more recent data from Bonn for $\pi^+$ production off
the proton \cite{Men77,Bue94,Zuc95}, and older Frascati \cite{Car73} and more
recent TRIUMF data
\cite{Bag88} on $\pi^{-}$ production off the neutron. As
shown in Fig.~\ref{fig:legs}, the predicted cross sections are in
perfect agreement
with the ratio $d\sigma_{\parallel}/d\sigma_{\perp}$ measured by
the LEGS collaboration \cite{Bla92,San96} whose data have not been included
in our fit. In  Fig.~\ref{fig:e_over_m} we show our result for the ratio
$R_{EM}$ which is in general agreement with the analysis of the
Virginia group \cite{Arn95}.

\section{The resonance pole parameters as determined by the speed plot}

The analytic continuation of a resonant partial wave as function of energy
into the second Riemann
sheet should generally lead to a pole in the lower half-plane. A
pronounced narrow peak reflects a time-delay in the scattering process
due to the existence of an unstable excited state. This time-delay is
related to the speed $SP$ of the scattering amplitude $T$, defined by
\cite{Hoe92,Hoe93}
\begin{equation}
SP(W) = \left\vert \frac{dT(W)}{dW}\right\vert ,
\end{equation}
where $W$ is the total $c.m.$ energy. In the vicinity of the resonance pole, the
energy dependence of the full amplitude $T = T_{B} + T_{R}$ is
determined by the resonance contribution,
\begin{equation}\label{T_res}
T_{R} (W) = \frac{r\Gamma_{R} e^{i\phi}}{M_{R}- W- i\Gamma_{R}/2}\,\,,
\end{equation}
while the background contribution $T_B$ should be a smooth function of
energy, ideally a constant. We note in particular that
$W_R = M_{R}- i\Gamma_{R}/2$ indicates the position of the
resonance pole in the complex plane, i.e. $M_{R}$ and $\Gamma_{R}$ are
constants and differ from the energy-dependent widths, and possibly
masses, derived
from fitting certain resonance shapes to the data\cite{RPP94}.
If the energy dependence of
$T_{B}$ is negligible, the speed is
\begin{equation}
SP(W) = r\Gamma_{R}\frac{\{[(M_{R}-W)^{2}-\Gamma_{R}^{2}/4]^2
          +\Gamma_{R}^2(M_{R}-W)^{2}\}^\frac{1}{2}}
          {\{(M_{R}-W)^{2}+\Gamma_{R}^{2}/4\}^{2}}.
\end{equation}
Obviously the speed has its maximum at $W = M_R$,
$SP(M_{R}) = 4r/\Gamma_{R} = H$,
and the half-maximum values are
$SP(M_{R} \pm \Gamma_{R}/2) = H/2$.
This determines the parameters $M_{R}$ and $\Gamma_{R}$ as well as the
absolute value $r$ of the residue. The
phase $\phi$ of the complex residue at the pole may be determined from an
Argand plot of the speed vector $dT/dW$.

\begin{figure}[htbp]
\centerline{\psfig{figure=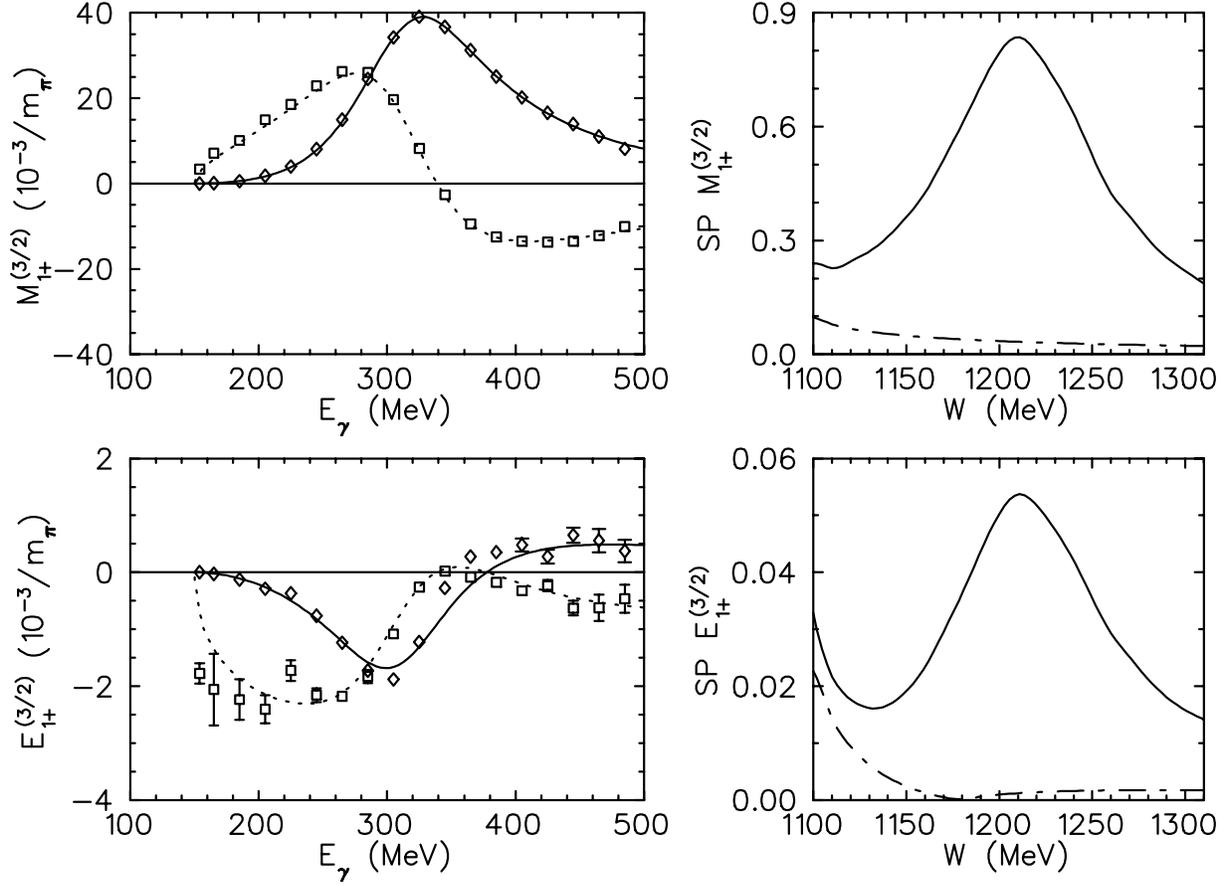,width=16cm,angle=90}}
\vspace{0.5cm}
    \caption{Our results for real (dotted lines) and imaginary (solid lines)
            parts of $M_{1+}^{(\frac{3}{2})}$ and $E_{1+}^{(\frac{3}{2})}$
            together with the speed plots of the full amplitudes. The
            dashed-dotted lines are the
            derivatives of the (nucleon) pole term contributions.
            The data are from the VPI analysis \protect\cite{Arn95}.}
  \label{fig:multi_sp}
\end{figure}

\begin{figure}[htbp]
\centerline{\psfig{figure=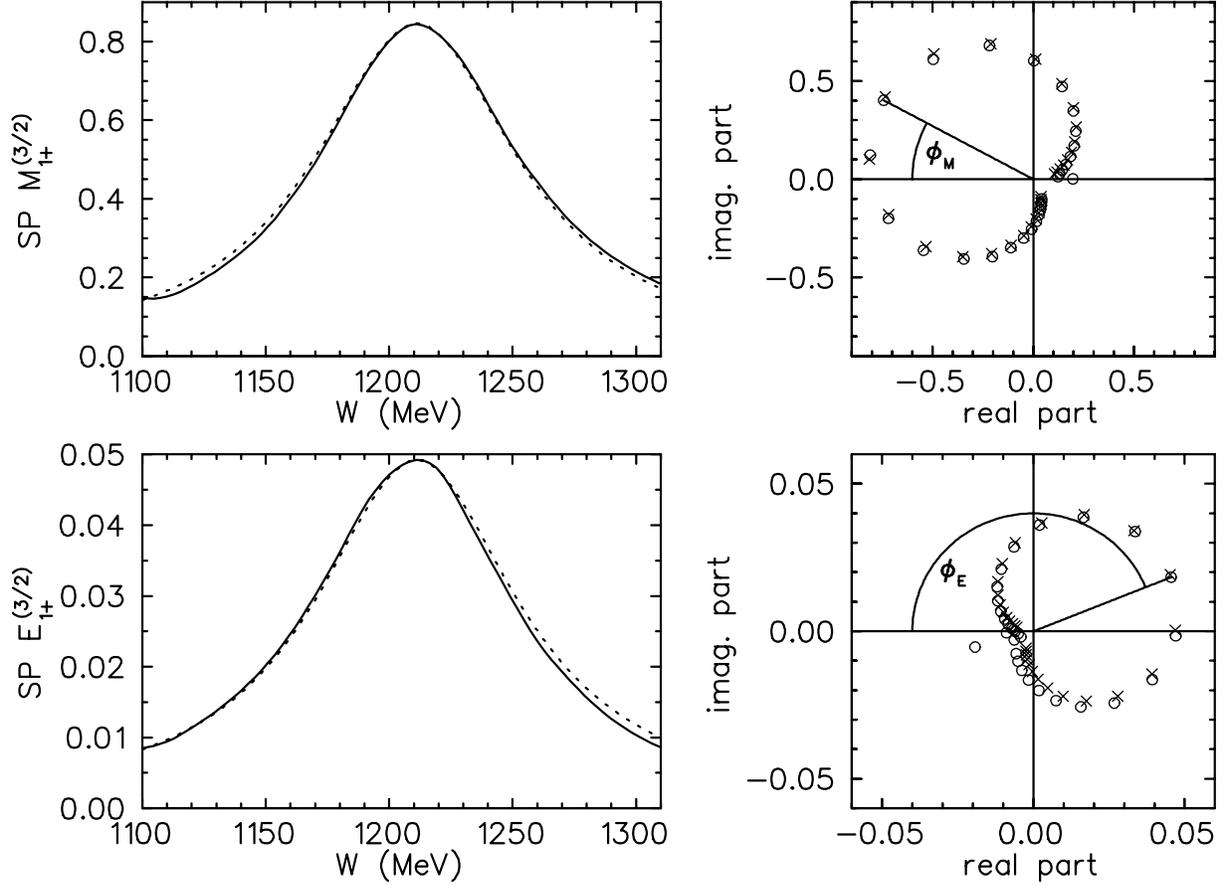,width=16cm,angle=90}}
\vspace{0.5cm}
    \caption{Speed of $M_{1+}^{(\frac{3}{2})}$ and $E_{1+}^{(\frac{3}{2})}$
     after subtraction of the (nucleon) pole term contributions (solid lines).
     For comparison we show the speed of an ideal resonance pole with
     the parameters given in Table~\protect\ref{tab:res_par}
     (dotted lines). In the Argand diagrams
     of the speed vectors we also compare our result (circles) with the
     speed corresponding to an exact pole (crosses). The only discrepancies are
     due to threshold effects. }
  \label{fig:sp_arg}
\end{figure}

\begin{figure}[htbp]
\centerline{\psfig{figure=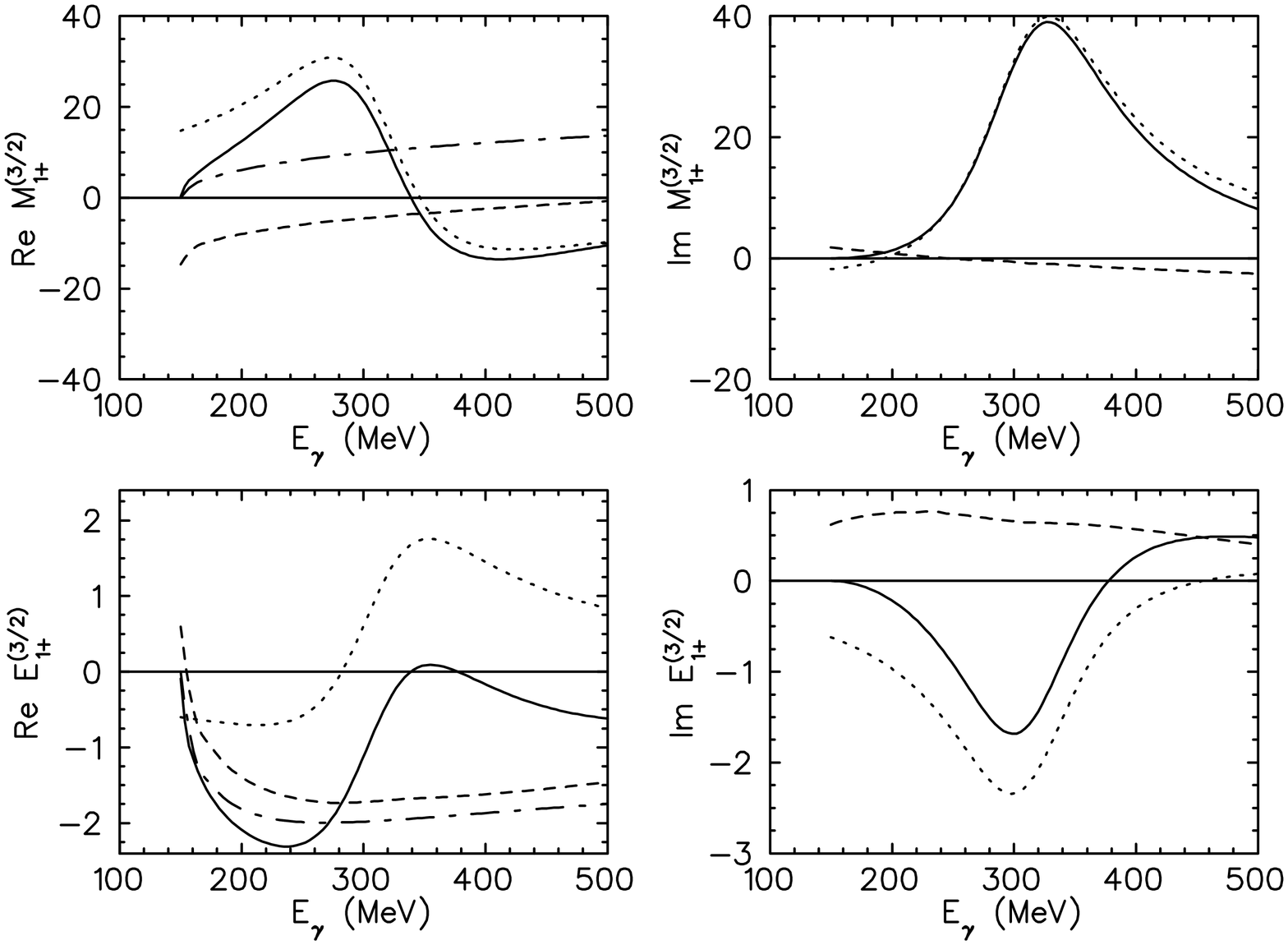,width=16cm,angle=90}}
\vspace{0.5cm}
    \caption{Separation of resonance and background contributions
     for the 3,3-multipoles.
      Solid lines: full amplitude,
      dotted lines: contribution of resonance pole,
      dashed lines: background,
      dash-dotted lines: (nucleon) pole term contribution.}
  \label{fig:uni_m}
\end{figure}

It should be noted that the speed plot technique requires a reasonably
smooth representation of the amplitude in order to differentiate it
in a meaningful way. As has been shown by H\"ohler \cite{Hoe92,Hoe93}
in the case of $\pi N$ scattering, dispersion relations are particularly
well suited for this purpose. If we apply the method to the partial
waves obtained by solving Eqs.~(\ref{inteq}), the results
for the $\Delta$ multipole clearly show a resonant peak, but the asymmetry
with respect to the maximum indicates an energy dependence of the background
(see Fig.~\ref{fig:multi_sp}).
This effect may be traced back to the nucleon pole terms. After subtracting
these well-defined terms from the amplitudes, the speed of both $E_{1+}$ and
$M_{1+}$ can be well described according to Eq.~(\ref{T_res}).
Fig.~\ref{fig:sp_arg}
shows a comparison
of this procedure to the ideal shape of a resonance pole and the
resulting Argand diagrams for the speed vector. Except for the threshold
region, the differences are almost invisible. Having determined all the
resonance parameters,
we can now decompose the full amplitudes into contributions of the resonance
pole and background terms. As may be seen in
Fig.~\ref{fig:uni_m}, the background
is a relatively smooth function of energy without any structure around the
resonance. However, the background is quite large, in particular in the
case of the $E_{1+}$ amplitude.

The resonance parameters derived from our analysis are shown in
Table~\ref{tab:res_par}. It
is seen that the pole position $W_{R} = M_{R} - i \Gamma_{R}/2
= (1211 - 50 i)$ MeV is in excellent agreement with the results obtained
from $\pi N$ scattering, $M_{R}= (1210\pm1)$ MeV and $\Gamma_{R} = 100$ MeV
\cite{Hoe92,Hoe93,RPP94} . This agreement may not
be very surprising in the case of the largely  resonant $M_{1+}$ amplitude,
for which there exist earlier investigations to determine the pole
position \cite{Cam76,Mir79}. However, it is much less obvious that
the interference pattern of
Re $E_{1+}$ of Fig.~\ref{fig:uni_m} should lead to the same answer.
The excellent
agreement in that case, too, is indeed very satisfactory and shows that
the speed plot
technique is quite reliable for the extraction of resonance properties. The
table also shows the absolute values and the phases of the resonance residues.
Because of the different backgrounds in the two amplitudes, $\phi_M$ and
$\phi_E$ are  different, and the ratio of the resonance amplitudes  is
complex. The fact that the two ''apple shaped'' structures in
Fig.~\ref{fig:sp_arg} are
essentially oriented in opposite direction is, however, related to the
negative value of $R_{EM}$ for the full (experimental) amplitude.
Concerning the
resonance pole contributions alone, we obtain
\begin{equation}
R_{\Delta} = \frac{r_{E} e^{i\phi^{E}}}{r_{M} e^{i\phi_{M}}}
= - 0.035 - 0.046 i.
\end{equation}

The ratio of the heights of the speed plots is $H_{E}/H_{M} =
r_{E}/r_{M} = 5.8\%$. We hasten to add, however, that the experimental
observable is related to the real part of the ratio (see Eq.~\ref{R_EM}), i.e.
the (unphysical) case of the resonance without background would lead to
$R_{EM}^{\mbox{\scriptsize res}}=$Re$(R_{\Delta}) = - 3.5\%.$

As has been mentioned before, the ratio for the full
(experimental) amplitudes is real below two-pion threshold due to the
Watson theorem. As may be seen from Fig.~\ref{fig:e_over_m}, this
ratio $R_{EM}(W)$ is strongly dependent on energy, and increases
with energy from negative
to positive values, e.g. $R_{EM}
(M_{R} - \Gamma_{R}/2) = -10.4\%$, $R_{EM} (M_{R}) = -4.3\%$, $R_{EM} (M_{R}
+\Gamma_{R}/2) =0.1\%$.
The resonance pole in the complex plane, $M_R - i \Gamma_{R}/2$, and the
nonresonant background lead to a $\pi N$ phase shift
$\delta_{1+} = 90^{\circ}$ at $W = M_{\Delta} = 1232$ MeV. Due to the
Watson theorem, both $E_{1+}^{(3/2)}$ and $M_{1+}^{(3/2)}$ are
completely  imaginary at this point, and the ratio can be determined
from the experimental data as
$R_{EM}(M_{\Delta})=$Im$ E_{1+}^{(3/2)} (M_{\Delta})/$Im$M_{1+}^{(3/2)}
(M_{\Delta})$.
The recent, nearly model-independent value of the Mainz group at
$W = M_{\Delta}$ is $(-2.5 \pm 0.2)\%$ \cite{Bec95,Kra96}.

\section{Conclusion}

It has been shown that the method of speed plots can be well applied to
analyze the pion photoproduction amplitudes. The resonance pole
position of the $\Delta (1232)$ is obtained from these amplitudes
in excellent agreement with the results from pion-nucleon
scattering. The complex residues of the resonance pole terms
give information on those parts of the full amplitude that have
a resonance--like behaviour. Whether such a contribution
originates from a ''bare'' resonance or, e.g. from a
nonresonant pion production followed by rescattering into
a resonant state, is model-dependent \cite{Wil96} and cannot be answered
by an analysis of the data but only within the framework of a specific
model.

In the future it will be interesting to analyze double polarization
variables, e.g. both photon and recoil or target polarization, because
some of these observables turn out to be very sensitive to electric
quadrupole radiation, too.
It is also worthwhile pointing out that reactions
like $e\vec{p} \rightarrow e'p\pi^{\circ}$ yield a longitudinal-transverse
interference term (''fifth structure function''), which is
sensitive to the imaginary part of the interference between the resonant
and background multipoles.

\acknowledgments

We would like to thank Prof.~G.~H\"ohler for very fruitful discussions and
the members of the A2 collaboration at Mainz for providing us with their
preliminary data, in particular R.~Beck, F.~H\"arter and H.-P.~Krahn.
This work was supported by the Deutsche Forschungsgemeinschaft (SFB~201).

\begin{table}[htbp]
    \caption{Resonance pole parameters determined by applying
     of the speed plot technique to our results for $E_{1+}^{(\frac{3}{2})}$
     and $M_{1+}^{(\frac{3}{2})}$.}
  \begin{center}
    \leavevmode
    \begin{tabular}{lrrrr}
     & $r$ $[10^{-3}\mathrm{MeV}/m_{\pi}]$ & $\phi$ [$^{\circ}$]
     & $M_{R}$ [MeV] & $\Gamma_{R}$ [MeV] \\
    \hline
    E & 1.23 & -154.7 & 1211$\pm$1 & 102$\pm$2\\
    M & 21.16 & -27.5 & 1212$\pm$1 & 99$\pm$2 \\
    \end{tabular}
    \label{tab:res_par}
  \end{center}
\end{table}

\vspace*{2cm}

\vspace*{2cm}

\newpage
\vspace*{2cm}

\newpage
\vspace*{2cm}

\newpage
\vspace*{2cm}


\begin{thebibliography}{99}
\bibitem{Bec65}C.~M.~Becchi and G.~Morpurgo, Phys. Lett. 17 (1965) 352.
\bibitem{Kon80}R.~Koniuk and N.~Isgur, Phys. Rev. D 21 (1980) 1868.
\bibitem{Ger81}S.~S.~Gershteyn and G.~V.~Dzhikiya, 
               Sov. J. Nucl. Phys. 34 (1981) 870.
\bibitem{Isg82}N.~Isgur, G.~Karl and R.~Koniuk, Phys. Rev. D 25 (1982) 2394.
\bibitem{Dre84}D.~Drechsel and M.~M.~Giannini, Phys. Lett. 143 B (1984) 329.
\bibitem{Bie87}J.~Bienkowska, Z.~Dziembowski and H.~J.~Weber,
               Phys. Rev. Lett. 59 (1987) 624.
\bibitem{Cap90}S.~Capstick and G.~Karl, Phys. Rev. D 41 (1990) 2767.
\bibitem{Cap92}S.~Capstick, Phys. Rev. D 46 (1992) 2864.
\bibitem{Kae83}G.~K\"albermann and J.~J.~Eisenberg, Phys. Rev. D 28 (1983) 71
               and D 29 (1984) 517.
\bibitem{Ber88}K.~Bermuth, D.~Drechsel, L.~Tiator and J.~B.~Seaborn,
               Phys. Rev. D 37 (1988) 89.
\bibitem{Wir87}A.~Wirzba and W.~Weise, Phys. Lett. B 188 (1987) 6.
\bibitem{Lei92}D.~B.~Leinweber, T.~Draper and R.~Woloshyn,
               Contribution to Baryons '92, p.~29 (1992).
\bibitem{Ols74}M.~G.~Olsson, Nucl. Phys. B 78 (1974) 55.
\bibitem{Koc84}J.~H.~Koch, E.~J.~Moniz and N.~Ohtsuka, Ann. Phys. (N.~Y.)
               154 (1984) 99.
\bibitem{Lag88}J.~M.~Laget, Nucl. Phys. A 481 (1988) 765.
\bibitem{Noz90}S.~Nozawa, B.~Blankleider and T.--S.~Lee,
               Nucl. Phys. A 513 (1990) 513.
\bibitem{Dav86}R.~M.~Davidson, N.~C.~Mukhopadhyay and R.~Wittman,
               Phys. Rev. Lett. 56 (1986) 804 and Phys. Rev. D 43 (1991) 71.
\bibitem{Kha93}M.~A.~Khandaker et al., $\pi N$ Newsletter 8 (1993) 114.
\bibitem{Ber93}A.~M.~Bernstein, S.~Nozawa and M.~A.~Moinester,
               Phys. Rev. C 47 (1993) 1274.
\bibitem{Van95}M.~Vanderhaeghen, K.~Heyde, J.~Ryckebusch, and M.~Waroquier,
               preprint SSF95-05-01, Gent (1995).
\bibitem{Sur95}Y.~Surya and F.~Gross, preprint CEBAF-TH-95-04 (1995).
\bibitem{Bla92}G.~S.~Blanpied et al., Phys. Rev. Lett. 69 (1992) 1880.
\bibitem{Bec95}R.~Beck, Proc. Int. Conf. ``Baryons '95'', Santa F\'e, (1995).
\bibitem{Kra96}H.--P.~Krahn, Ph.D. thesis, Mainz (1996).
\bibitem{Wat54}K.~M.~Watson, Phys. Rev. 95 (1954) 228.
\bibitem{Sch69}D.~Schwela and R.~Weizel, Z. Physik 221 (1969) 71.
\bibitem{Pfe72}W.~Pfeil and D.~Schwela, Nucl. Phys. B 45 (1972) 379.
\bibitem{Fuc96}M.~Fuchs et al., Phys. Lett. B 368 (1996) 20.
\bibitem{Hae96}F.~H\"arter, PhD. thesis, Mainz (1996).
\bibitem{Men77} D.~Menze, W.~Pfeil and R.~Wilcke, Compilation of pion
                photoproduction data, Bonn (1977).
\bibitem{Bue94}K.~Buechler et al., Nucl. Phys. A 570 (1994) 580.
\bibitem{Zuc95}H.~Dutz, PhD.~thesis, Bonn (1993),\\
               D.~Kr\"amer, PhD.~thesis, Bonn (1993),\\
               B.~Zucht, PhD.~thesis, Bonn (1995).
\bibitem{Car73}F.~Carbonara et al., Nuovo Cim. 13 A (1973) 59.
\bibitem{Bag88}A.~Bagheri et al., Phys. Rev. C 38 (1988) 875.
\bibitem{San96}A.~Sandorfi, private communication (1996), release L7a8.0.
\bibitem{Arn95}R.~A.~Arndt, I.~I.~Strakovsky and R.~L.~Workman,
               Phys. Rev. C 53 (1996) 430.
\bibitem{Hoe92}G.~H\"ohler and A.~Schulte, $\pi N$ Newsletter 7 (1992) 94.
\bibitem{Hoe93}G.~H\"ohler, $\pi N$ Newsletter 9 (1993) 1.
\bibitem{RPP94}Review of Particle Properties, Phys. Rev. D 50 (1994).
\bibitem{Cam76}R.~R.~Campbell, G.~L.~Shaw and J.~S.~Ball,
               Phys. Rev. D 14 (1976) 2431.
\bibitem{Mir79}I.~I.~Miroshnichenko, V.~I.~Nikiforov, V.~M.~Sanin,
               P.~V.~Sorokin, and S.~V.~Shalatskii,
               Sov. J. Nucl. Phys. 29 (1979) 94.
\bibitem{Wil96}P.~Wilhelm, Th.~Wilbois and H.~Arenh\"ovel,
            preprint~MKPH-T-96-1, Mainz (1996).
\end{thebibliography}
\end{document}